\documentclass[sigconf,natbib,screen=true]{acmart}
\usepackage{booktabs} % For formal tables
\usepackage[skip=0pt]{caption}
\usepackage{multirow}
\usepackage{bm}
\usepackage{balance}
\usepackage{algorithm}
\usepackage{algorithmic}
\usepackage{lineno}
\usepackage[inline]{enumitem}
\usepackage{acronym}
\usepackage{pifont}
\usepackage{array}
\usepackage{graphicx}
\usepackage{subcaption}
\copyrightyear{2020}
\acmYear{2020}
\setcopyright{rightsretained}

\acmYear{2020}
\copyrightyear{2020}
\setcopyright{acmcopyright}
\acmConference{AIIS'20}{July 30, 2020}{Virtual Event, China}
\acmPrice{TBA}
\acmDOI{TBA}
\acmISBN{TBA}

\AtBeginDocument{%
  \providecommand\BibTeX{{%
    \normalfont B\kern-0.5em{\scshape i\kern-0.25em b}\kern-0.8em\TeX}}}
\acrodef{RS}{Recommender System}
\acrodef{SR}{Sequential Recommender}
\acrodef{SRec}{Sequential Recommendation}
\acrodef{ADSR}{Attribute-aware Diversifying Sequential Recommender}
\acrodef{AE}{Attribute-aware Encoder}
\acrodef{AP}{Attribute Predictor}
\acrodef{ADD}{Attribute-aware Diversifying Decoder}
\acrodef{BSR}{Base Sequential Recommender}
\acrodef{ANAM}{Attribute-aware Neural Attentitive Model}
\acrodef{MTASR}{Multi Task Attribute-aware Sequential Recommender}
\acrodef{MMR}{Maximal Marginal Relevance}

\DeclareMathOperator{\softmax}{softmax}

\author{Anton Steenvoorden$^{1,2}$\qquad
Emanuele Di Gloria$^{2}$\qquad
Wanyu Chen$^{1, 3}$
}
\author{Pengjie Ren$^{1}$\qquad
Maarten de Rijke$^{1,2}$}
\makeatletter
\def\authornotetext#1{
\if@ACM@anonymous\else
    \g@addto@macro\@authornotes{
    \stepcounter{footnote}\footnotetext{#1}}
\fi}
\makeatother
\affiliation{
 \institution{
 \textsuperscript{\rm 1}University of Amsterdam, Amsterdam, The Netherlands. 
 \textsuperscript{\rm 2}Ahold Delhaize, Zaandam, The Netherlands.
 }
 \institution{
 \textsuperscript{\rm 3}National University of Defense Technology, Changsha, China
 }
}

\email{{anton.steenvoorden, emanuele.di.gloria}@aholddelhaize.com, {w.chen2, p.ren, m.derijke}@uva.nl}
\def\authors{Anton Steenvoorden, Emanuele Di Gloria, Wanyu Chen, Pengjie Ren, and Maarten de Rijke}

\acmSubmissionID{1}

\begin{document}

\title{Attribute-aware Diversification for Sequential Recommendations}

\renewcommand{\shortauthors}{Steenvoorden et al.}

\begin{abstract}
Users prefer diverse recommendations over homogeneous ones. 
However, most previous work on \aclp{SR} does not consider diversity, and strives for maximum accuracy, resulting in homogeneous recommendations. 
In this paper, we consider both accuracy and diversity by presenting an \acf{ADSR}. 
Specifically, \ac{ADSR} utilizes available attribute information when modeling a user's sequential behavior to simultaneously learn the user's most likely item to interact with, and their preference of attributes. 
Then, \ac{ADSR} diversifies the recommended items based on the predicted preference for certain attributes.
Experiments on two benchmark datasets demonstrate that \ac{ADSR} can effectively provide diverse recommendations while maintaining accuracy.
\end{abstract}

%%
%% The code below is generated by the tool at http://dl.acm.org/ccs.cfm.
%% Please copy and paste the code instead of the example below.
%%



\maketitle

% !TEX root =  ./cikm2020_anton_recommendation.tex

\section{Introduction}
\acp{RS} are widely used to help users find items they are interested in.
Traditional \acp{RS} such as Collaborative Filtering and Matrix Factorization \cite{su2009survey} provide recommendations using a decomposition of a user-item interaction matrix.
These approaches do not take the order of interactions into account and fail to capture the users' evolving preferences.
\acfp{SR} have attracted a lot of attention, as they can exploit the order of interactions \cite{repeatnet}.
In industry, \acp{SR} allow e-commerce retailers to provide recommendations to users, based on their sequence of interactions and have proven to be very effective \cite{repeatnet, li2017neural, hidasi2016session}.

Side information in \acp{SR}, e.g., item attributes (category, genre, etc.), has proven to be useful for capturing user preferences. 
E.g., \citet{bai2018attribute} use available item attributes to make their model attribute-aware by obtaining a unified representation from items and their attributes. 
And \citet{chen2019air} use item attributes to infer users' future intention, allowing them to do intent-aware recommendation. 
However, the focus of these methods lies on increasing performance in terms of accuracy only, resulting in homogeneous recommendations.
Moreover, diversity is an important metric to consider as it has been shown that users prefer diverse search results opposed to highly accurate,
but redundant ones~\cite{zhang2008avoiding, carbonell1998use}.

We propose to use item-attribute information to diversify \acp{SR}.
For instance, if we know that a user enjoys \textit{documentaries} and \textit{thrillers}, and this user is in search of films about ``Artificial Intelligence'' (AI), they are likely interested in both a \textit{documentary} about AI and in a \textit{thriller} where general AI achieves world domination.
Genre information and preferences can be used to present diverse recommendations related to AI, collectively covering multiple genres. 
To this end, we present an \acf{ADSR}, which considers both accuracy and diversity while generating the list of recommendations.

The \ac{ADSR} consists of three modules: an \ac{AE}, an \ac{AP}, and an \ac{ADD}.
The \ac{AE} models the sequence of item interactions and the sequence of item-attributes. 
The \ac{AP} learns and predicts the user's preference on attribute level.
Finally, the \ac{ADD} incrementally generates diversified recommendations by using the predictions from the \ac{AP} and trading off accuracy and diversity.

The \ac{AE} and \ac{AP} both make predictions and are therefore optimized in a multi-task learning paradigm~\cite{caruana1997multitask}.
We carry out experiments on two benchmark datasets.
The results demonstrate that \ac{ADSR} benefits from modeling and predicting item attributes, and can provide attribute-aware diversified recommendations while preserving accuracy.

To sum up, the contributions of this work are as follows:
\begin{itemize}[leftmargin=*,nosep]
\item We propose \ac{ADSR}, which is one of the first to address diverse recommendations for \acp{SR}.
\item We devise \ac{AP} and \ac{ADD} modules to generate attribute-aware diversified recommendations by jointly optimizing item recommendation and item attribute prediction as an auxiliary task.
\end{itemize}

% !TEX root =  ./sigir2020_anton_recommendation.tex

% \section{Related Work}

% \subsection{Sequential recommendation}

% \subsection{Attribute-aware recommendation}

% \subsection{Diversified recommendation}
% !TEX root =  ./cikm2020_anton_recommendation.tex

\section{\acf{ADSR}}
Given user $u$ with behavior sequence $S_v = \{v_1, \ldots, v_t, \ldots, v_T\}$ and corresponding item attribute sequence $S_c = \{c_1, \ldots, c_t, \ldots, c_T\}$, where $v_t$ is the item $u$ interacts with at time step $t$, and $c_t$ is the attribute of $v_t$, we aim to create a diversified list of recommendations $R_L$.
Formally, let $P(c_j \mid S_c, S_v)$ be the importance of $c_j$ based on the sequences and let $P(R_L \mid S_v, S_c, c_j)$ be the probability of recommending $R_L$ conditioned that $c_j$ is the user's preferred attribute.
Then, we find $R_L$ by maximizing $P(R_L \mid S_v, S_c)$, defined as:
\begin{equation}
    P(R_L \mid S_v, S_c) = \sum_{j=1}^{|C|} P(R_L \mid S_v, S_c, c_j) \hspace{1pt} P(c_j \mid S_c, S_v),
\end{equation}

Optimising $P(R_L \mid S_v, S_c)$ directly is difficult due to the large search space~\cite{agrawal2009diversifying}.
Therefore, we propose to generate $R_L$ iteratively, by appending the item with the highest score $S(v_i)$ to $R_L$ at each time step, similar to \citep{carbonell1998use, agrawal2009diversifying}:
\begin{equation}
\label{eq:s_v}
    S(v_i) = \lambda_{\text{s}} \cdot S_\text{rel}(v_i) + (1-\lambda_{\text{s}}) \cdot S_\text{div}(v_i),
\end{equation}
where $S_\text{rel}(v_i)$ is the relevance score for item $v_i$ (see \S\ref{sec:item_encoder}), and $S_\text{div}(v_i)$ is the attribute-aware diversity score (see \S\ref{sec:AAD}). The hyperparameter
$\lambda_\text{s}$ is used to balance $S_\text{rel}$ and $S_\text{div}$, giving control over the accuracy and diversity of the trained model.

\begin{figure}
    \centering
    \includegraphics[width=0.99\columnwidth]{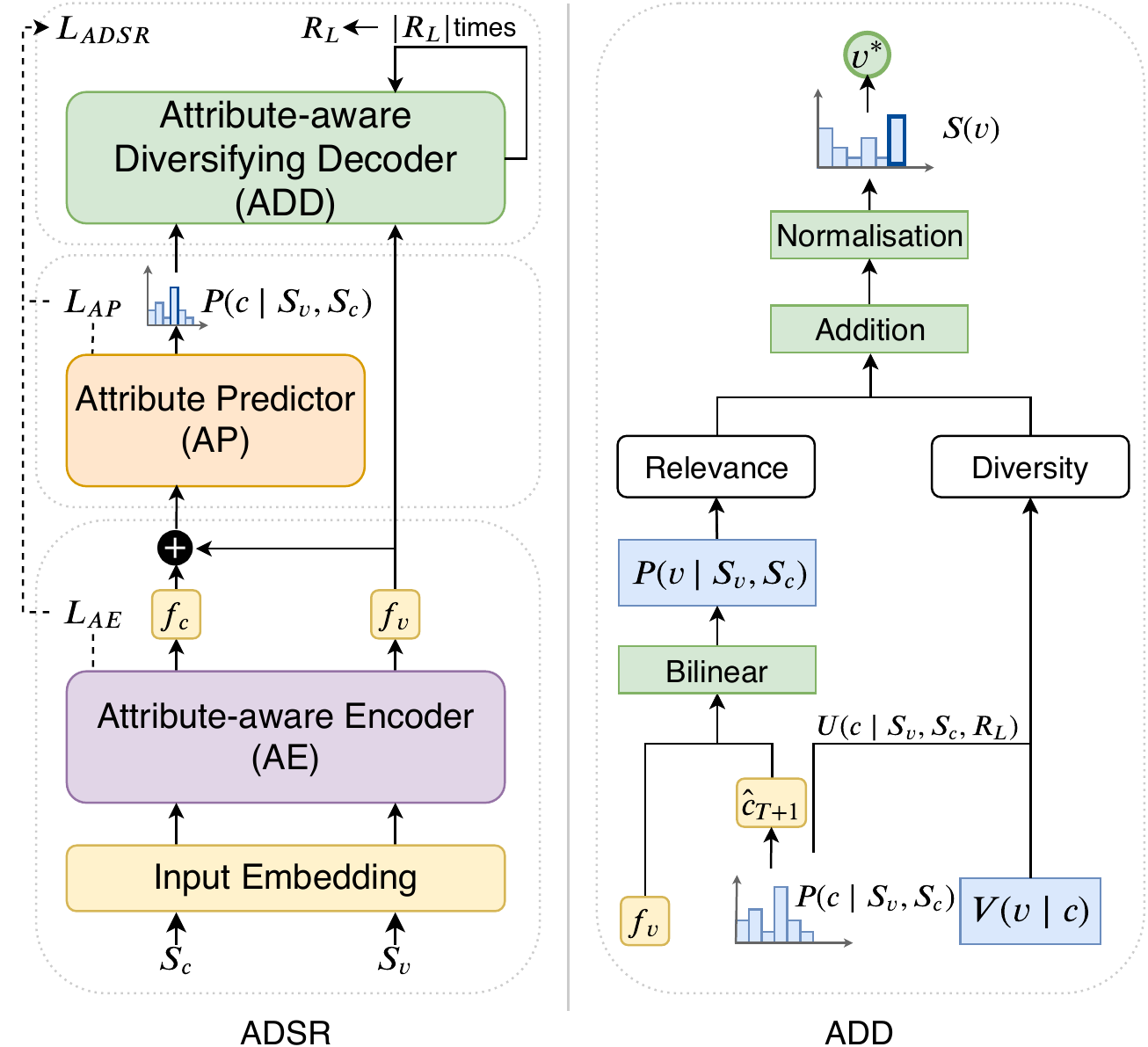}
    \caption{\acf{ADSR} overview. The input to the recommender are item and attribute sequences. The output is a diversified list of recommendations based on learned attribute preferences.}
    \vspace*{-\baselineskip}
    \label{fig:proposed_architecture}
\end{figure}

We propose \ac{ADSR} to model $S_\text{rel}(v_i)$ and $S_\text{div}(v_i)$, as shown in Figure \ref{fig:proposed_architecture}. 
The \acf{AE} models the input sequences $S_c, S_v$ to get $\mathbf{h}_{c}$ and $\mathbf{h}_{v}$, respectively. 
Next, the \acf{AP} predicts the next item attribute distribution $P(\mathbf{c} \mid S_v, S_c)$. 
Finally, the \acf{ADD} uses these outputs to generate the diversified list $R_L$.

\subsection{\acl{AE}}\label{sec:item_encoder}
First, the sequence of attributes is encoded using a bidirectional RNN with GRU-cells~\cite{li2017neural}. 
The input of this GRU is a sequence of item attribute embeddings concatenated with the item embeddings: $ \{\mathbf{p}_1, \mathbf{p}_2, \ldots, \mathbf{p}_T\}$ where $\mathbf{p}_t = [\mathbf{c}_t; \mathbf{v}_t] \in \mathbb{R}^{d_c + d_v}$.
The output of the GRU are hidden representations $\{\mathbf{h}_{c_1}, \mathbf{h}_{c_2}, \ldots, \mathbf{h}_{c_T}\}$ where $\mathbf{h}_{c_t} \in \mathbb{R}^{2d_\text{GRU}}$.
Then, a second bidirectional RNN is used to get the representation for the sequence of item interactions. 
The input to this encoder is the concatenation of the item embeddings and the hidden states: $\mathbf{q}_t = [\mathbf{v}_t ; \mathbf{h}_{c_t}]$, yielding $\{\mathbf{h}_{v_1}, \mathbf{h}_{v_2}, \ldots, \mathbf{h}_{v_T}\} \in \mathbb{R}^{2d_\text{GRU}}$. 

Next, we employ additive attention to get the global preference $\mathbf{f}_{v}$ and $\mathbf{f}_{c}$, capturing the complete sequences, as follows: 
\begin{equation}\label{eq:attention}
\begin{split}
\mathbf{f}_{v} &= {} \sum_{j=1}^T \alpha_{Tj} \mathbf{h}_{v_t} \\
\alpha_{Tj} &= {} \softmax \left(\mathbf{W}_p  \tanh(\mathbf{W}_q \mathbf{h}_{v_T} + \mathbf{W}_k \mathbf{h}_{v_j}) \right),
\end{split}
\end{equation}
where matrices $\mathbf{W}_q, \mathbf{W}_k$ are used to transform $\mathbf{h}_i$ into a latent space, and $\mathbf{W}_p \in \mathbb{R}^{2d_\text{GRU} \times 1}$ is used to get the attention weights.
$\mathbf{f}_{c}$ is calculated as in Eq.~\ref{eq:attention} using the attention weights $\alpha_{Tj}$ obtained for $\mathbf{f}_{v}$.
Finally, \ac{AE} calculates the relevance score as:
\begin{equation}
\label{eq:s_rel}
S_\text{rel}(v_i) = P(v_i \mid S_v, S_c) = 
% \softmax(\mathbf{g}^T\mathbf{v}_i)
\frac{\exp{(\mathbf{g}^T \mathbf{v}_i})}{\sum_{j=1}^{ |V|} \exp{(\mathbf{g}^T \mathbf{v_j}})},
\end{equation}
where $\mathbf{g} = [\mathbf{f}_{v}; \mathbf{\hat{c}}_{T+1}] \mathbf{W}_g \in \mathbb{R}^{d_v}$ and  ${\mathbf{\hat{c}}_{T+1} = \sum_{j=1}^{|C|} P(c_j \mid S_v, S_c) \cdot \mathbf{c}_j}$ is the weighted average item attribute based on \ac{AP}'s predicted attribute preference.
% The loss induced by \ac{AE} ($L_AE$) is described in Sec. \ref{sec:losses}.
For clarity, the loss induced by \ac{AE} ($L_{AE}$) is described in Section \ref{sec:losses}.
% The loss calculated for predictions made by \ac{AE} is the Cross-Entropy Loss: 
% \begin{equation}
% \begin{split}
% L_{AE} = \frac{-1}{|V|} \sum_{i=1}^{|V|} y_i \log P(v_i \mid S_v, S_c),\\
% y_i=1 \text{ if } v_i = v_{T+1}, \text{ else } 0  \text { and } v_{T+1} \text{ is the target}
% \end{split}
% \end{equation} 

\subsection{\acl{AP}}
The \ac{AP} is of great importance to our model.
The output ${P(\bold{c} {\mid} S_v, S_c)}$ is used by the \ac{AE} to determine item relevance, and by the \acf{ADD} (Sec. \ref{sec:AAD}) to diversify the recommendations.
The \ac{AP} predicts $P(\bold{c} \hspace{1pt}|\hspace{1pt} S_v, S_c)$ with a small neural network. 
The input is the concatenation of the global preference: $[\mathbf{f}_{v};\mathbf{f}_{c}]$.
In doing so, the \ac{AP} exploits features from both encoders when making the prediction.
The \ac{AP} consists of a hidden layer, with Batch Normalisation and ReLU activation, followed by the output layer activated with a sigmoid function.
This prediction task is an auxiliary task, with an additional loss $\text{L}_{AP}$ (see Section \ref{sec:losses}). 

\subsection{\acl{ADD}}\label{sec:AAD}
The \ac{ADD} is responsible for generating the diversified list of recommendations $R_L$. 
Inspired by IA-Select, a method used in web search \cite{agrawal2009diversifying}, we incrementally extend $R_L$ by selecting at each step the item with the highest score (Eq.~\ref{eq:s_v}), i.e., $v_i^* = \operatorname*{arg\,max}_{v_i} S(v_i)$.
Specifically, at each step the \ac{ADD} calculates $S_{\text{div}}$ using $P(\bold{c} \mid S_v, S_c)$, obtained from the \ac{AP}, as initial estimation of importance per category $U$:
\begin{equation}\label{eq:s_div}
S_\text{div}(v_i) = \sum_{j=1}^{|C|} U(c_j \mid S_v, S_c, R_L) \hspace{1pt}  \left(1 - V(v_i \mid c_j) \right),
\end{equation}
where $V(v_i \mid c_j)$ represents the value of $v_i$ in context of $c_j$, and is $1$ if $v_i$ belongs to category $c_i$ and $0$ else.
After each step, we update $U(c_j \mid S_v, S_c, R_L)$ to reflect the newly added item to $R_L$ by:
\begin{equation}
\begin{split}
    U&(c_j {\mid} S_v, S_c, R_{L}\cup \{v_i^*\}) = {}\\
      & \softmax\left[\left(1 {-} V(v_i^* \mid c_j) \right) U(c_j {\mid} S_v, S_c, R_L)\right].
\end{split}    
\end{equation} 
% The \ac{ADD} is only used during inference. 

\subsection{Losses}\label{sec:losses}
The loss $L_{AE}$ calculated for the \ac{AE} is the Cross-Entropy Loss. 
\begin{equation}
\begin{split}
L_{AE} & = \frac{-1}{|V|} \sum_{i=1}^{|V|} y_i \log P(v_i \mid S_v, S_c)\\
y_i & =1 \text{ if } v_i = v_{T+1}, \text{ else } 0  \text {, where } v_{T+1} \text{ is the target.}
\end{split}
\end{equation} 
To calculate $L_\text{AP}$, the Binary Cross-Entropy Loss is used, allowing for multi-labeled targets:
\begin{equation}\label{eq:l_adsr}
\begin{split}
\mbox{}\hspace*{-2mm}
L_{AP} & = \frac{-1}{|C|}\sum_{i=1}^{|C|} y_i \log P(c_i {\mid} S_v, S_c) +(1{-}y_i) \log \left(1 - P(c_i {\mid} S_v, S_c) \right) 
\hspace*{-2mm}\mbox{}\\
% L_{AP} &= \frac{-1}{|C|}\sum_{i=1}^{|C|} y_i \log P(c_i | S_v, S_c) + (1-y_i) \log \left( 1- P(c_i | S_v, S_c) \right),\\
y_i &=1 \text{ if } c_i = c_{T+1}, \text{ else } 0 \text {, where } c_{T+1} \text{ is the target.}
\end{split}
\end{equation}
Then, to form the final loss used to optimize \ac{ADSR} ($L_{ADSR}$), the two losses are interpolated with balancing parameter $\lambda_{\text{MT}}$: 
\begin{equation}
{L_{ADSR} = \lambda_{\text{MT}} \cdot L_{AE} + (1{-}\lambda_{\text{MT}}) \cdot L_{AP}}.
\end{equation}

% \subsection{Losses}\label{sec:losses}
% The loss calculated for the \ac{AE} is the Cross-Entropy Loss:
% \begin{equation}
% \begin{split}
% L_{AE} & = \frac{-1}{|V|} \sum_{i=1}^{|V|} y_i \log P(v_i \mid S_v, S_c)\\
% y_i & =1 \text{ if } v_i = v_{T+1}, \text{ else } 0  \text {, where } v_{T+1} \text{ is the target.}
% \end{split}
% \end{equation} 
% To calculate $L_\text{AP}$, the Binary Cross-Entropy Loss is used instead, allowing for multi-labeled targets. 
% Then, to form the final loss used to optimize \ac{ADSR} ($L_{ADSR}$), the two losses are interpolated with balancing parameter $\lambda_{\text{MT}}$.
% That is, ${L_{ADSR} \hspace{1pt}{=}\hspace{2pt} \lambda_{\text{MT}} * L_{AE} + (1{-}\lambda_{\text{MT}}) * L_{AP}}$, where
% \begin{equation}\label{eq:l_adsr}
% \begin{split}    
% L_{AP} & = \frac{-1}{|C|}\sum_{i=1}^{|C|} y_i \log P(c_i \mid S_v, S_c) +{}\\
% &\mbox{}\hspace*{3cm} (1-y_i) \log \left( 1- P(c_i \mid S_v, S_c) \right),\\
% % L_{AP} &= \frac{-1}{|C|}\sum_{i=1}^{|C|} y_i \log P(c_i | S_v, S_c) + (1-y_i) \log \left( 1- P(c_i | S_v, S_c) \right),\\
% y_i &=1 \text{ if } c_i = c_{T+1}, \text{ else } 0 \text {, where } c_{T+1} \text{ is the target.}
% \end{split}
% \end{equation}

% !TEX root =  ./cikm2020_anton_recommendation.tex

\section{Experimental Setup}
We seek to answer the following questions in our experiments:
\begin{enumerate*}[label=(RQ\arabic*), leftmargin=*]
\item \label{rq:1} Where does the improvement of \ac{ADSR} come from? What are the effects of the \ac{AE}, \ac{AP} and \ac{ADD} modules? (See \S\ref{s:as}.)
\item \label{rq:2} What is the performance of \ac{ADSR} compared with non-attribute-aware diversity methods? (See \S\ref{s:op}.)
\item \label{rq:3} How does the trade-off parameter $\lambda_\text{s}$ affect the performance of \ac{ADSR}? (See \S\ref{s:effect_lambda}.)
% \item \label{rq:4} Can \ac{ADSR} generate proper diversified recommendations improving user experience? What are the limitations? (See \S\ref{s:cs}.)
\end{enumerate*}

\subsection{Datasets}
We evaluate \ac{ADSR} on the following real-world datasets:
% \begin{itemize}[leftmargin=*,nosep]
    % \item MovieLens-1M,\footnote{\label{foot:datasets}\url{https://grouplens.org/datasets/movielens/1m}} which contains 1 million ratings from 6,000 users over 4,000 movies. Movies belong to at least one of 18 genres.
    % \item TMall,\footnote{\url{https://tianchi.aliyun.com/dataset/dataDetail?dataId=53}} which is an e-commerce dataset containing 44,528,127 user interactions over 2,353,207 items belonging to one of 72 categories. We use ``buy'' interactions only.
% \end{itemize}
\begin{itemize}[leftmargin=*,nosep]
    \item MovieLens-1M,\footnote{\label{foot:datasets}\url{https://grouplens.org/datasets/movielens/1m}} which contains 1 million ratings from 6K users over 4K movies. Movies belong to at least one of 18 genres.
    \item TMall,\footnote{\url{https://tianchi.aliyun.com/dataset/dataDetail?dataId=53}} which is an e-commerce dataset containing 44.5M user interactions over 2.4M items belonging to one of 72 categories. We use the ``buy'' interactions only.
\end{itemize}
Genre and category are used as item attribute.
% Genre and category are used as item attribute for ML1M and TMall respectively.
Multi-labeled genres are handled by summing the embeddings.\footnote{In this work, a single genre/category is used as attribute, but \ac{ADSR} can be modified to use other/multiple attributes (or contexts) using e.g. embedding summation.}
Both datasets are filtered from users and items with less than 20 interactions.
Per sequence, the first 80\% is used as training set. 
The 20\% is filtered from sequences with unseen items and halved to form the validation and test sets.
The final inputs are sliding windows of size 10~\cite{hidasi2016session} (10th item as target).
Statistics after processing are reported in Table \ref{tab:data-stats}.

% \begin{table}[]
% \caption{Statistics per dataset after pre-processing}
% \label{tab:data-stats}
% % \resizebox{.95\columnwidth}{\height}
% % {
% \begin{tabular}{@{}lccc@{}}
% \toprule
% Dataset                 & ML1M & TMall \\ \midrule
% \# Users                &  6,041 & 31,855\\
% \# Items                & 3,261 & 58,344\\
% \# Training Sequences   & 784,309 & 698,081\\
% \# Validation Sequences & 93,929 & 54,706 \\
% \# Test Sequences       & 97,871 & 54,705\\
% \# Total Sequences      & 976,109 & 807,492\\
% \# Item Attributes      & 18   & 71    \\ \bottomrule
% \end{tabular}
% % }
% \end{table}

\begin{table}[]
\caption{Dataset statistics after pre-processing.}
\label{tab:data-stats}
\setlength{\tabcolsep}{2.5pt}
\resizebox{\columnwidth}{!}{%
\begin{tabular}{@{}lcccccc@{}}
\toprule
 & \#Users & \#Items & \#Train seq. & \#Valid seq. & \#Test seq. & \#Attributes \\ \midrule
ML1M & \phantom{3}6,041 & \phantom{3}3,261 & 784,309 & 93,929 & 97,871 & 18 \\
TMall & 31,855 & 58,344 & 698,081 & 54,706 & 54,705 & 71 \\ \bottomrule
\end{tabular}%
}
\end{table}

\subsection{Evaluation metrics}
Following \cite{li2017neural, hidasi2016session, repeatnet}, we measure accuracy of the model using \textit{MRR@k} and \textit{Recall@k}.
Predictions made by the \ac{AP} for multi-labeled targets are considered correct if one of the active labels is predicted.
% \textit{Recall} is a measure that explains how often the test item is presented in the recommended list. 
% \textit{MRR} measures the ranking accuracy of the recommender, placing the test item at the first position of the list results in high MRR.
To measure diversity, we use \textit{Intra-List Diversity (ILD)}~\citep{zhang2008avoiding}:
\begin{equation}\label{eq:dissimilarity}
ILD(R_L) = \frac{2}{|R_L|\cdot(|R_L|-1)} \sum_{i \in R_L} \sum_{j \neq i \in R_L} d(i, j),
\end{equation} 
where $d(i,j)$ is the euclidean distance between the one-hot-encoded item attributes of $v_i$ and $v_j$. 
A second diversity measure is used where the number of unique attributes in the recommended list is counted, we refer to this as \textit{discrete diversity} (\textit{Dis}).

\subsection{Models for comparison}
A number of \ac{SR} methods have been proposed in the last few years.
We do not compare with them because: 
\begin{enumerate*}[label=(\arabic*)] 
\item improving recommendation accuracy is out of the scope of this paper; 
and \item many are not directly comparable due to the use of different model architectures.
\end{enumerate*}
Therefore, we construct/select baselines that are fair (use the same information, similar architectures, etc.) to compare with:
\begin{itemize}[leftmargin=*,nosep]
\item \textit{\ac{BSR}} uses only $S_v$ to generate hidden representations. Then, \ac{BSR} applies additive attention to get the global preference, used to generate $R_L$, similar to NARM~\cite{li2017neural}. 
% \item \textit{\ac{BSR}} uses only $S_v$ to generate hidden representations, and applies an additive attention mechanism to get global preference $\bold{f}_{v_T}$, directly used to generate $R_L$, similar to NARM \cite{li2017neural}.
\item \textit{\ac{ANAM}} incorporates attribute information and applies attention to the hidden representations, similar to \ac{ANAM} \cite{bai2018attribute}. We have adapted it to do \ac{SR} rather than next-basket recommendation, and modified it to be similar to our own variants. The core principles remain the same, both apply attention to a unified representation based on $S_v, S_c$.
\item \textit{\ac{MTASR}} ex-\newline tends \ac{ANAM} to predict the attribute preference, and to use the weighted attribute embedding to predict $R_L$. This model can be regarded as a special case of \ac{ADSR} with $\lambda_\text{s}=1$.
\item \textit{\ac{ANAM}+MMR} is \ac{ANAM} with the \acl{MMR} reranking algorithm by \citet{carbonell1998use}, trading off relevance and diversity by selecting $v^*$ by: 
\begin{equation}
v^* = \operatorname{arg\,max}_{v_i \in V \backslash R_L} \lambda S(v_i) + (1-\lambda) \min_{v_j \in R_L} d(i,j),
\end{equation}
where V are all items not in $R_L$ and $d(i,j)$ is the same as in Eq.~\ref{eq:dissimilarity}. This model does not have the \ac{AP} module, so it cannot do attribute-aware diversification based on preference.
\end{itemize}
Besides \ac{ANAM}+\ac{MMR}, none of the methods listed considers diversity when recommending items. 
\ac{ADSR} learns attribute preferences to provide an attribute-aware diversified list of recommendations.

\subsection{Implementation details}
For a fair comparison, we use the same settings for all models.
% We set ${d_v{=}d_c{=}d_\text{GRU}{=}128}$.
We set ${d_v=d_c=d_\text{GRU}=128}$.
Embeddings are initialized using the Xavier method. 
Dropout is applied to the embeddings and hidden representations with $p=0.5$.
We use the Adam optimizer with learning rate ${0.01}$, batches of 1024 samples for 60 epochs.
% The final hidden representation is projected to $d_v$ using a single linear layer in the case of \textit{\ac{BSR}} and \textit{\ac{ANAM}}.
$\bold{f}_{v}$ is projected to $d_v$ by one linear layer in the case of \textit{\ac{BSR}} and \textit{\ac{ANAM}}.
% A bilinear mapping is used to combine and project the hidden representations in \textit{\ac{MTASR}} and \textit{\ac{ADSR}}.
A bilinear mapping is used to combine and project $\bold{f}_{v}$ and $\bold{f}_{c}$ in \textit{\ac{MTASR}} and \textit{\ac{ADSR}}.
The best model is selected by MRR and Recall on the validation set.%
\footnote{The code from this paper is available at \href{https://github.com/antonsteenvoorden/ADSR}{https://github.com/antonsteenvoorden/ADSR}.}
% !TEX root =  ./cikm2020_anton_recommendation.tex
% \input{05-figs}

\newcommand{\und}[1]{\underline{#1} }

\begin{table*}[]
\caption{Performance of recommendation models. The best performing model is boldfaced, the second best is underlined.
% Our model is significantly better, determined by a two-tailed $t$-test
Statistical significance is determined by a two-tailed $t$-test. Pairwise differences of all models vs. \textit{\ac{BSR}} are significant with $p \leq .01$. 
For \textit{\ac{ADSR}} vs. \textit{\ac{ANAM}+\ac{MMR}} significance with $ p \leq .05$ is indicated by $^\vartriangle$, the other results are significant with $p \leq .01$. 
% The values of the hyperparameters $\lambda_\text{s}$ and $\lambda_\text{MT}$ are denoted beneath the dataset name.
%Significance is indicated by $^\blacktriangle$ and $^\vartriangle$ for $p \leq .01$ and $ p \leq .05$ respectively. 
}
\label{tab:results}
\setlength{\tabcolsep}{2.5pt}
\resizebox{\textwidth}{!}{%
\begin{tabular}{@{\extracolsep{3pt}}l cc cc cc cc c cc cc cc cc c @{}}
\toprule
& \multicolumn{9}{c}{\textbf{ML1M}} & \multicolumn{9}{c}{\textbf{TMall}} \\ 
& \multicolumn{9}{c}{\ac{ANAM}+\ac{MMR}: $\lambda_\text{s}@10=0.2, \lambda_\text{s}@20=0.3$} &  \multicolumn{9}{c}{\ac{ANAM}+\ac{MMR}:$\lambda_\text{s}@10=0.01, \lambda_\text{s}@20=0.01$}\\
& \multicolumn{9}{c}{\ac{ADSR}: $\lambda_\text{s}@10=0.25, \lambda_\text{s}@20=0.4, \lambda_\text{MT}=0.9$} &  \multicolumn{9}{c}{\ac{ADSR}: $\lambda_\text{s}@10=0.01, \lambda_\text{s}@20=0.01, \lambda_\text{MT}=0.4$}\\
\cmidrule{2-10} \cmidrule{11-19}
Model & \multicolumn{2}{c}{MRR} & \multicolumn{2}{c}{Recall} & \multicolumn{2}{c}{ILD} & \multicolumn{2}{c}{Dis.} &
\ac{AP} Acc. &
\multicolumn{2}{c}{MRR} & \multicolumn{2}{c}{Recall} & \multicolumn{2}{c}{ILD} & \multicolumn{2}{c}{Dis.} &
\ac{AP} Acc.\\ 
 \cmidrule{2-3} \cmidrule{4-5} \cmidrule{6-7} \cmidrule{8-9} \cmidrule{10-10} \cmidrule{11-12} \cmidrule{13-14} \cmidrule{15-16} \cmidrule{17-18} \cmidrule{19-19} 
 & @10 & @20 & @10 & @20 & @10 & @20 & @10 & @20 & & @10 & @20 & @10 & @20 & @10 & @20 & @10 & @20 & \\ \midrule
\ac{BSR} & 0.0728 & 0.0793 & 0.0728 & 0.2805 & 1.4197 & 1.452 & 7.7593 & 10.2603& -- & 0.0863 & 0.0891 & 0.1601 & 0.2013 & 0.9141 & 0.9428 & 5.0294 & 7.9399 & -- \\
\ac{ANAM} & 0.0777 & 0.0840 & 0.1944 & 0.2871 & 1.4283 & 1.4601 & 7.8751 & 10.4070 & -- & 0.0983 & 0.1015 & \und{0.1835} & \und{0.2298} & 0.8374 & 0.8725 & 4.6215 & 7.3883 & -- \\
\ac{MTASR} & \textbf{0.0804} & \textbf{0.0868} & \textbf{0.2000} & \textbf{0.2936} & 1.4250 & 1.4512 & 7.9025 & 10.4311 & 0.5292 &
\textbf{0.0984} & \textbf{0.1018} & \textbf{0.1848} & \textbf{0.2332} & 0.7680 & 0.8069 & 4.2306 & 6.7749 & 0.3820 \\
\midrule
\ac{ANAM}+\ac{MMR} & 0.0772 & 0.0839 & 0.1920 & 0.2852 & \und{1.5367} & \und{1.5016} & \und{9.9375} & \und{11.9467} & -- & \und{0.0978} & \und{0.1007} & 0.1808 & 0.2235 & \und{0.9417} & \und{1.0095} & \und{5.9315} & \und{11.8666} & -- \\
% MTASR+MMR &  &  &  &  &  &  &  &  &  &  &  &  &  &  &  &  \\
\midrule
\ac{ADSR} & \und{0.0795} & \und{0.0862} & \und{0.1939} & \und{0.2859} & \textbf{1.6222} & \textbf{1.5025} & \textbf{10.7988} & \textbf{12.7599} & 0.5292 & 
0.0963$^\vartriangle$ & 0.0989 & 0.1744 & 0.2108 & \textbf{1.0267} & \textbf{1.1275} & \textbf{6.5234} & \textbf{13.8204} & 0.3820 \\ \bottomrule
\end{tabular}%
}
\end{table*}

\section{Experimental results}

\subsection{Ablation study}
\label{s:as}
To answer \ref{rq:1} and to show the effectiveness of the \ac{AE}, \ac{AP} and \ac{ADD} modules, we compare the results of variations of \ac{ADSR}, as shown in Table~\ref{tab:results}.
\ac{BSR} is \ac{ADSR} without all three modules (no attribute information is used).
Further, \ac{ANAM} is \ac{ADSR} without \ac{AP} and \ac{ADD}, and \ac{MTASR} is \ac{ADSR} without \ac{ADD}.

First, the \ac{AE} is effective.
This is demonstrated by the fact that \ac{ANAM} outperforms \ac{BSR} on both datasets, showing that modeling attribute information significantly improves performance.
Second, the \ac{AP} further enhances performance of the model, as \ac{MTASR} outperforms \ac{ANAM}. 
This shows that learning attribute preference with the \ac{AP} brings a significant improvement to the model.
% However, this is not the case on TMall, \ac{MTASR} is slightly worse than \ac{ANAM} on MMR and Recall@10 (Recall@20 did improve).
% We think the reason is that the \ac{AP} is limited by the sparsity of this dataset. The increased size of the attribute space further amplifies this.
%
Third, the \ac{ADD} significantly improves diversity without hurting accuracy.
This claim is supported by the fact that \ac{ADSR} outperforms \ac{MTASR} on both diversity metrics. 
\ac{MTASR} outperforms \ac{ADSR} in terms of MRR and Recall, which is expected as diversity is increased by trading off relevance, where \ac{MTASR} is equal to \ac{ADSR} with $\lambda_\text{s}{=}1$.
However, the trade-off is advantageous, as gains in diversity are made of 39\% and 14\% over \ac{BSR} in ILD@10 and Dis@10, with a reduction of merely 1.24\% in MRR@10 and 3.29\% in Recall@10.

\subsection{Comparison with non-attribute-aware diversity methods}
% \subsection{Comparison with diversity methods}
\label{s:op}
To answer \ref{rq:2}, we compare \ac{ADSR} with \ac{ANAM}+\ac{MMR}.
By incorporating \ac{MMR} into \ac{ANAM}, \ac{ANAM}+\ac{MMR} is able to provide diversified recommendations.
From Table~\ref{tab:results} we see that \ac{ADSR} is able to provide more diverse recommendations on both datasets.
On ML1M, \ac{ADSR} significantly outperforms \ac{ANAM}+\ac{MMR} on all metrics, while increasing diversity: \ac{ADSR} is more effective at diversifying results, with an increase of $8.67\%$ and $5.56\%$ over \ac{ANAM}+\ac{MMR} in ILD@10 and Dis@10, while yielding higher MRR and Recall.
On TMall, \ac{ADSR} is best at diversifying results, showing increases of 9.98\% and 9.03\% over \ac{ANAM}+\ac{MMR} in terms of Dis@10 and ILD@10, and gains of 16.46\% and 11.69\% in Dis@20 and ILD@20. 
However, \ac{ADSR} yields lower accuracy than \ac{ANAM} and \ac{ANAM}+\ac{MMR}, 
which is likely due to the high sparsity of TMall and the larger attribute space, making it harder for the \ac{AP} to correctly predict attributes.
% The reported numbers favor diversity, trading in more accuracy for large increases in diversity. 
% However, for $\lambda_s{=}0.02@20$, \ac{ADSR} is able to closely match performance while slightly increasing diversity.
% \todo{this value is not reported anywhere, is this ok?}

\subsection{Effect of the trade-off hyperparameter $\lambda_\text{s}$}\label{s:effect_lambda}
\begin{figure}
    \centering
    \begin{subfigure}[t]{0.5\columnwidth}
        \centering
        \includegraphics[height=3cm, trim={0 0 1.45cm 1.35cm}, clip]{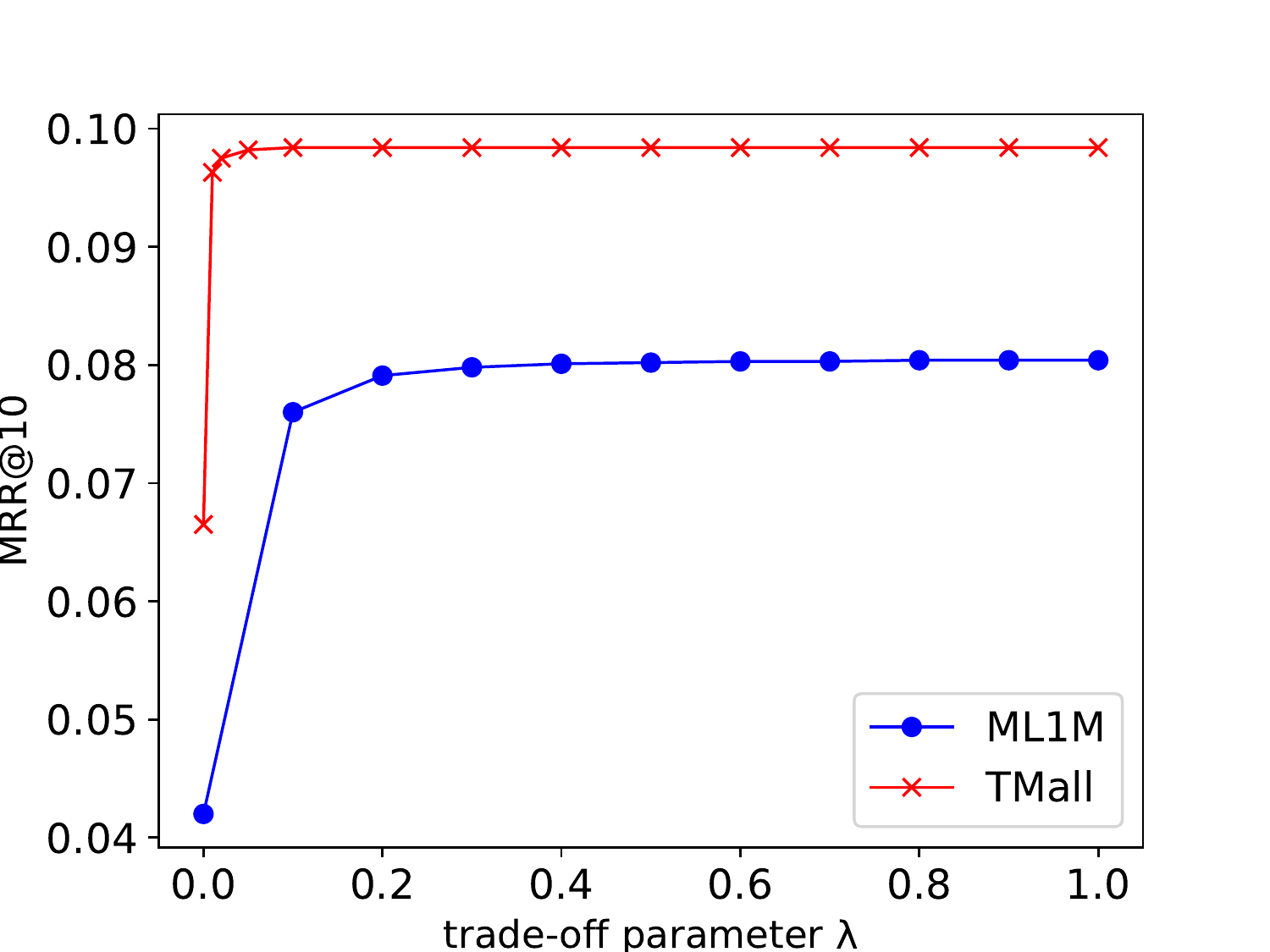}
        % \caption{Lorem ipsum}
    \end{subfigure}%
    \begin{subfigure}[t]{0.5\columnwidth}
        \centering
        \includegraphics[height=3cm, trim={0 0 1.45cm 1.35cm}, clip]{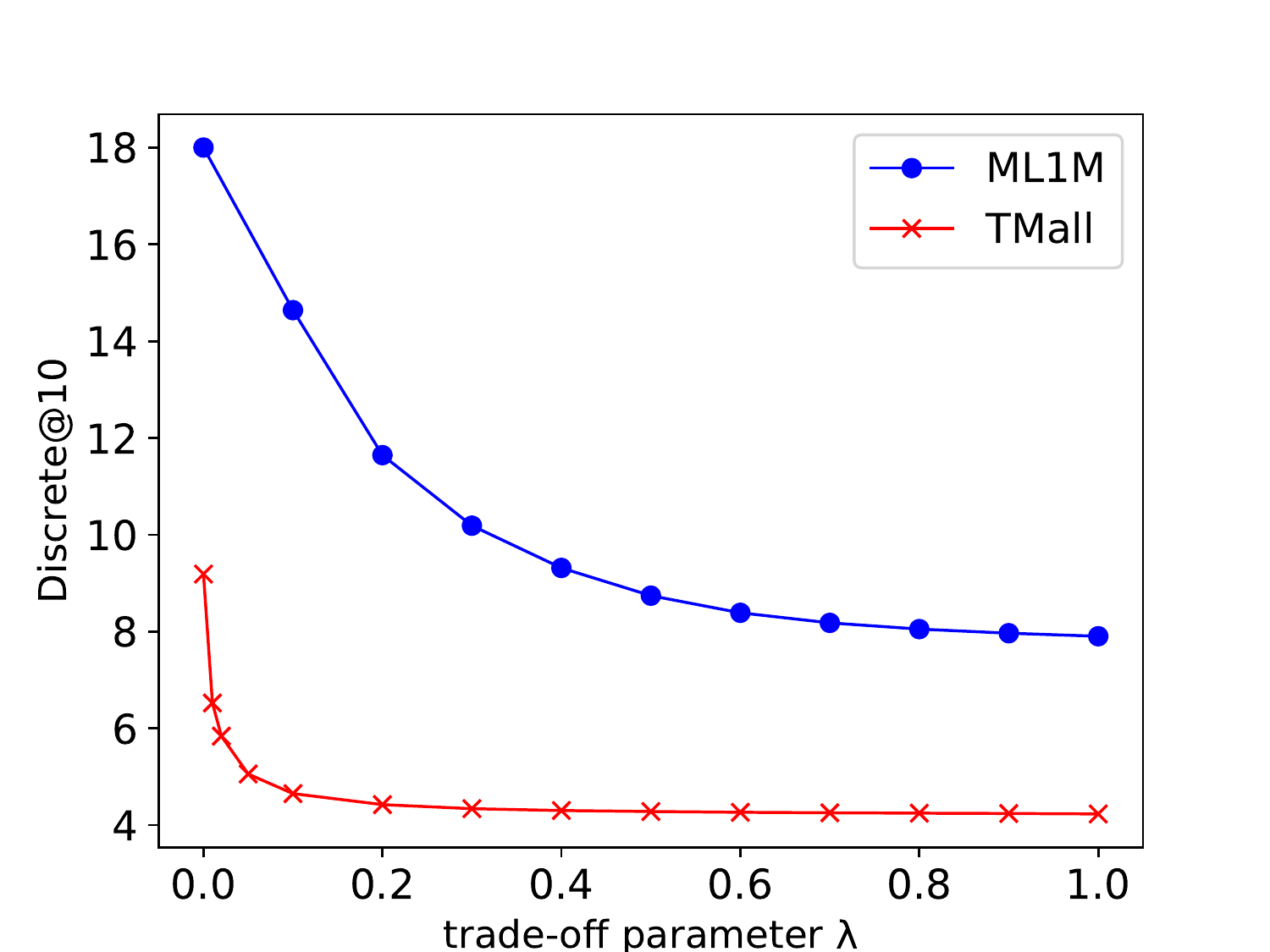}
        % \caption{test}
    \end{subfigure}
    \caption{Performance of \ac{ADSR} in \textit{MRR@10} and \textit{Dis@10}, while varying $\lambda_\text{s}$ (in Eq. \ref{eq:s_v}) from $0.0$ to $1.0$ on two datasets.}
    \vspace*{-\baselineskip}
    \label{fig:varying_lambda}
\end{figure}

To answer \ref{rq:3}, we vary $\lambda_\text{s}$ from $0.0$ to $1.0$ and investigate the effect it has on accuracy and diversity (for brevity, we chose a single metric of each type; see Figure \ref{fig:varying_lambda}). 
When $\lambda_\text{s}{=}0$, it maximizes diversity only, and does this successfully, yielding the maximum diversity possible on both datasets.
On TMall, very low values of $\lambda_\text{s}$ are required to diversify results as the item space is much larger, resulting in very small values for $S_\text{div}$. 
Clearly, $S_\text{div}$ only affects item selection when it is weighed much heavier than $S_\text{rel}$.
Finally, Figure \ref{fig:varying_lambda} shows that diversity decays slower than accuracy increases, meaning that \ac{ADSR} can effectively achieve a large gain in diversity for a small loss of accuracy.

% \subsection{Case study}
% \label{s:cs}
% \todo{can IA select allows multiple items from same category?, if strong preference}
%To asnwer \ref{rq:4}
% !TEX root = ./cikm2020_anton_recommendation.tex

\section{Conclusion and Future Work}
We propose a novel \acl{ADSR} that utilizes attribute information when modeling a user's sequential behavior to simultaneously learn their most likely item to interact with and their preference for attributes.
Experimental results on two datasets show
that \ac{ADSR} effectively diversifies the list of recommendations based on the predicted preference for attributes, trading in a controllable and small amount of accuracy for large gains in diversity, through hyperparameter $\lambda_\text{s}$. 
Further, ablation shows the importance and the positive effect of incorporating attribute information and attribute prediction.

A limitation of \ac{ADSR} is that performance decreased on TMall, due to sparsity and a larger item and attribute space. 
However, dealing with sparsity is a common issue in \ac{SR}.
As to future work, we would like to address the limitation of the \ac{AP} by looking into ways of leveraging auxiliary information such as user profiles, item reviews, etc. as we think high \ac{AP} accuracy increases performance.
% \todo{Another limitation is the inconsistency of $\lambda_s$, which needs to be tuned per dataset, this might be addressed by normalizing $S_\text{div}$}

% \section*{Code and data}
% The code used to run the experiments in this paper is available at \url{http: //url.suppressed.for.anonymity}.

% \begin{acks}
% To Robert, for the bagels and explaining CMYK and color spaces.
% \end{acks}

\bibliographystyle{ACM-Reference-Format}
\bibliography{bibtex}
% \bibliography{bibtex_maarten}

\end{document}